\documentclass[12pt]{article}
\usepackage[dvips]{graphics}
\usepackage{amssymb}
\usepackage[utf8]{inputenc}
\usepackage{xcolor}
\usepackage{graphics}
\usepackage{epsfig} 
\usepackage{tikz}
\usepackage[export]{adjustbox}
\usepackage{animate}
\usepackage{amsmath}
\usepackage{cite}
\usepackage{hyperref}
\usepackage{verbatim}
\usetikzlibrary{arrows,shapes}
\usepackage{times}
\usetikzlibrary{calc}
\newcommand{\be}{ \begin{equation} }
	\newcommand{\ee}{\end{equation}} 
\begin{document} 
	\def\theequation{\arabic{section}.\arabic{equation}} 
	
		\title{	Embedding the stationary spacetimes into  Brans-Dicke cosmology via conformal transformations}
		
		\author{Dilek KAZICI ÇİFTCİ$^{a}$\footnote{email address: dkazici@nku.edu.tr} \ and  Ali DEMİRÖZLÜ\footnote{demirozluali@gmail.com} \\ \\ 
			{\small   }\\ 
			{\small   $^{a}$ Department of Physics, Tekirdağ Nam{\i}k Kemal University, }\\{\small $59030$, Süleymanpaşa/ Tekirda\u {g}, Turkey} 
		}  
		
		\date{} 
		\maketitle 
		\vspace*{1truecm}

		\vspace*{1truecm} 

\begin{abstract}
	A conformal transformation of a static or stationary spacetime by a time dependent conformal scale factor  ${S(\tau )}^2$ is one of the methods of producing a cosmological spacetime. Using this knowledge and Brans-Dicke (BD) field equations, we investigate two time dependent metrics, including Friedmann-Lemaitre-Robertson-Walker (FLRW) spacetime and conformally transformed Kerr-Newman black hole, and we obtain solutions that allow different expansion rates for each geometry. These expansion rates depend on the matter content of the conformally transformed geometry. We state that the BD scalar field yields accelerated expansion of the conformal spacetime if the original metric has vacuum geometry, and no acceleration if the original spacetime has some curvature or matter content in it. \newline
\end{abstract}

\section{Introduction}  

The expansion of  universe was observed by Hubble in 1920s and also recently discovered that the universe is not only expanding but also accelerating \cite{Riess,Perlmutter}. Observational evidence, coming from the type Ia Supernova explosion, implies that the correct spacetime geometry must be nonstatic both in astrophysics and cosmology. Since gravitational interactions can be well described in the General Theory of Relativity, to understand the structure and behavior of the universe, we use the General Relativistic formulations and pseudo-Riemannian geometry as a mathematical tool. On sub-galactic regions like the Solar system, the effects of gravity are not strong enough and spacetime can be characterized as nearly flat. In these scales, the General Theory of Relativity (GR) has been accurately tested and verified \cite{Bertotti,Reasenberg,Will1,Will2,Everitt,Berti}. However, when we use GR in cosmology especially at the large scale structures or for the evolving universe in time, we need some new phenomena which we have not understood and explained theoretically and observationally yet, for example, dark matter and dark energy \cite{Amendola1,Amendola2}. Also, black holes provide strong gravitational fields and there are large deviations from GR at high field strength \cite{Sakstein1,Yagi,Abuter,Burrage}. This means that, for a more general theory of gravitation, we need to understand the strong gravitational regimes and the large scale structure of the universe \cite{Koyama,Baker1,Clifton}. Brans-Dicke (BD) scalar tensor theory is a well known scenario of gravitational field \cite{Brans,Bergmann,Wagoner,Nordvedt}, and in general, it is considered as an alternative theory to GR. In our sense, this is not an alternative to GR but a more general theory of gravity and it can be related with the $f\mathrm{(}R\mathrm{)}$ theory \cite{Sotiriou,Felice,Nojiri}, string theory \cite{Callan,Fradkin,Blachke} and Kaluza Klein theory  \cite{Billyard,Overduin} in the appropriate limits. Also, BD theory involves Mach's principle which says that all of the matter in the universe affects each other, hence a universe, filled with a scalar field, might be a reasonable candidate for this interaction of the masses. Therefore, motivated to find the solutions for an accurate cosmological model and also a theory for highly gravitating cosmological environment, it might be convenient to study BD theory of gravity.

In general, the cosmological expansion in time has been defined by a time dependent scale factor in front of the spatial part of the metric components as in the FLRW metric. For more general cases, we do not restrict ourselves with the FLRW case, we can also include Kerr-Newman black hole. Using the method of conformal transformation by rescaling static or stationary spacetime, we try to produce a cosmological model for asymptotically non-flat cosmological black holes. In this context, conformal transformation of Kerr-Newman black hole also provide to obtain inhomogeneities in the FLRW backgrounds. BD scalar tensor theory adds the system a scalar degree of freedom which is represented by a scalar field $\phi $. Using this property, we find a relation between the BD scalar field $\phi $ and the conformal scale factor $S\mathrm{(}\tau \mathrm{)}$ in which the transformed spacetime has the stress-energy tensor for a perfect fluid. This means that the stress-energy tensor arises from the curvature and matter content of spacetime, and also that the expansion parameter is closely related to the scalar field $\phi$. Also, in reference \cite{Valerio4}, the author works on the scalar field and perfect fluid  and concludes that the scalar field and a perfect fluid are not equivalent but a convenient correspondence for the formal purposes.

A stationary/static submanifold can be embedded in a cosmological background by a scale factor of $S\mathrm{(}\tau \mathrm{)}$ as,
\begin{equation}
	\tilde{g}_{ab}=S(\tau)^2g_{ab}\label{transformation}.
\end{equation}
This expression is a conformal transformation of a metric tensor with a conformal factor $S(\tau )^2$. In the rest of the paper, the metric $g_{ab}$ will be called as original metric or submanifold $\mathcal{M}$ and chosen as static or stationary spacetime, and the transformed metric ${\tilde{g}}_{ab}$ will be named as conformally rescaled frame or cosmological background $\tilde{\mathcal{M}}$, which describes a spacetime evolving in time. Depending on the properties of the chosen original frame, the ${\tilde{g}}_{ab}$  may characterize a cosmological spacetime or a dynamical object. In the rest of the paper, all of the geometric quantities in this conformally rescaled cosmological frame will be denoted by a ``tilde''. In this work, we will first start with the vacuum case of original frame in which $G_{ab}\mathrm{=0}$. Next, we will consider other contents of matter sources such as $G_{ab}\mathrm{=}T^{EM}_{ab}$ and for the more general case $G_{ab}=T_{ab}$, where $T^{EM}_{ab}$ is Maxwell stress-energy tensor and $T_{ab}$ represents the energy-momentum tensor of a perfect fluid.

The conformal transformation rules (\ref{transformation}) in GR frame require Einstein tensor to be \cite{Wald, Carroll, Valerio book,Dabrowski},
\begin{eqnarray}
	G_{ab}&=&\tilde{G}_{ab}+3\tilde{g}_{ab}
	\frac{\tilde{\nabla}_cS\tilde{\nabla}^cS}{S^2}+\frac{2}{S}(\tilde{\nabla}_a\tilde{\nabla}_b S-\tilde{g}_{ab}\tilde{\square}S)\label{EFE},
\end{eqnarray}
where, $G_{ab}=R_{ab}-\frac{1}{2}g_{ab}R$ is the Einstein tensor for the static/stationary submanifold and $\tilde{G}_{ab}=\tilde{R}_{ab}-\frac{1}{2}\tilde{g}_{ab}\tilde{R}$ is the Einstein tensor for the conformally rescaled space-time, the covariant dervative $\tilde{\nabla}_a$ and d'Alembertian $\tilde{\square }$ are taken with respect to the metric ${\tilde{g}}_{ab}$.  This equation shows that even if there is no matter in the untilted manifold ($G_{ab}=0$), under the conformal transformation (\ref{transformation}), the transformed spacetime geometry may contain any form of matter which may be responsible for the expansion of the universe in time. In this work, we suppose that, the spacetime geometry ${\tilde{g}}_{ab}$, obtained by the conformal transformation (\ref{transformation}) of a static or stationary spacetime, would be a cosmological solution of Brans-Dicke theory in the Jordan frame and we can find a relation between the conformal factor $S(\tau )$ and the BD scalar scalar field $\phi$.  In other word, the scalar field would be responsible for the expansion of the spacetime. Hence, we will use BD scalar tensor theory, and its action in the conformally transformed frame as,
\begin{equation}
	S_{BD}= \int d^4x \sqrt{ -\tilde{g}} \left\{ \tilde{\phi} \tilde{R} 
	-\frac{\omega}{\tilde{\phi}}  \tilde{g}^{ab} \tilde{\nabla}_a \tilde{\phi} 
	\tilde{\nabla}_b \tilde{\phi} -
	\tilde{\mathcal{L}}_m \right\}\, , \label{BD action}
\end{equation} 
which also includes matter Lagrangian ${\tilde{\mathcal{L}}}_m$. Here, $\tilde{\phi }$ is called as BD scalar field might be a function of both time and spatial coordinates, and $\omega $ is a free dimensionless BD parameter. The variation of the BD action \eqref{BD action} with respect to ${\tilde{g}}^{ab}$ gives the field equations as,
\begin{eqnarray}
	&&\tilde{G}_{ab}-
	\frac{\omega }{\tilde{\phi}^2} \left( 
	\tilde{\nabla}_a \tilde{\phi} \tilde{\nabla}_b \tilde{\phi} -\frac{1}{2}\, \tilde{g}_{ab} \tilde{g}^{cd}\tilde{\nabla}_c 
	\tilde{\phi} \nabla_d \tilde{\phi}  \right) -\frac{1}{\tilde{\phi}}  \left( \tilde{\nabla}_a \tilde{\nabla}_b \tilde{\phi}  
	-\tilde{g}_{ab} \tilde{\square} \tilde{\phi} \right)\nonumber\\
	&& -\frac{8\pi}{\tilde{\phi} } \, \tilde{T}_{ab} =0
	\, ,  \label{BDfe}
\end{eqnarray}
and variation with respect to $\tilde{\phi }$ gives the scalar field equation as,
\begin{equation}
	\tilde{\square}\tilde{\phi}=\frac{8\pi }{2\omega+3}\tilde{T}.\label{Scalarfield}
\end{equation}
Using equations (\ref{EFE})   and (\ref{BDfe}), we can write the stress-energy tensor ${\tilde{T}}_{ab}$ in terms of the Einstein tensor of submanifold $\mathcal{M}$, scale factor $S(\tau )$ and scalar field $\tilde{\phi }$ as,
\begin{eqnarray}
	\frac{8\pi}{\tilde{\phi} } \, \tilde{T}_{ab} &=&G_{ab}-3\tilde{g}_{ab}
	\frac{\tilde{\nabla}_cS\tilde{\nabla}^cS}{S^2}-\frac{2}{S}(\tilde{\nabla}_a\tilde{\nabla}_b S-\tilde{g}_{ab}\tilde{\square}S)\nonumber\\
	\nonumber\\
	&&-
	\frac{\omega }{\tilde{\phi}^2} \left( 
	\tilde{\nabla}_a \tilde{\phi} \tilde{\nabla}_b \tilde{\phi} -\frac{1}{2}\, \tilde{g}_{ab} \tilde{\nabla}_c 
	\tilde{\phi} \nabla
	\tilde{\phi}  \right) -\frac{1}{\tilde{\phi}}  \left( \tilde{\nabla}_a \tilde{\nabla}_b \tilde{\phi}  
	-\tilde{g}_{ab} \tilde{\square} \tilde{\phi} \right) 
	\,, \label{SEtensor}
\end{eqnarray}
here, even if we  suppose $G_{ab}\mathrm{=0}$, there would be a nonzero stress energy tensor in the conformal frame. It means that the conformal transformation creates an extra term composed of the conformal factor, and this term can be related to the BD scalar field. 

The energy-momentum tensor of the matter in conformal frame is,
\begin{equation}
	{\tilde{T}}^{ab}_m=\frac{2}{\sqrt{-\tilde{g}}}\frac{\delta }{\delta {\tilde{g}}_{ab}}\left(\sqrt{-\tilde{g}}{\tilde{L}}_m\right), \label{variationT}
\end{equation}
and in the form of perfect fluid  of matter it is defined as,
\begin{eqnarray}
	\tilde{T}_{ab}=\tilde{T}_{ab}^{\,(pf)}=(\tilde{P}+\tilde{\rho})\tilde{u}_a\tilde{u}_b+\tilde{g}_{ab}\tilde{P},\label{perfectfluid}
\end{eqnarray}
the four velocity ${\tilde{u}}_a$ and ${\tilde{u}}_a{\tilde{u}}^a=-1$ , $\tilde{\rho }\ $and $\tilde{P}$ are energy density and pressure respectively. 

In BD theory, the matter part of Lagrangian ${\tilde{\mathcal{L}}}_m$ is not coupled with the scalar field $\tilde{\phi }$, this is the main difference between the $\mathrm{BD}$ and Jordan models. But as we see in \eqref{SEtensor}, the stress-energy tensor of the matter part seems to couple with the scalar $\tilde{\phi }$, however, it is not a coupling, but an interaction between the scalar field and metric tensor field. Therefore, the weak equivalence principle is respected \cite{Fujii}. To see this interaction explicitly, we begin by assuming an ansatz given by \cite{Johri},
\begin{equation}
	\tilde{\phi}=\phi_0 S(\tau)^\alpha \label{phi&S},
\end{equation}
with ${\tilde{\phi }}_0$ and $\alpha $ constants and ${\tilde{\phi }}_0\ge 0$. This equation yields BD scalar field depends only on time \cite{Akarsu}. In this work we suppose that the BD scalar field naturally arises in the universe and stating the ansatz (\ref{phi&S}), this scalar field may provide different expansion rates depending on the matter in the submanifold of the metric in (\ref{transformation}).

Field equations (\ref{SEtensor}) indicate that if the relation between scalar field and conformal factor becomes as in (\ref{phi&S}), each term with $S$ and $\tilde{\phi }$ on the right will be related with eachother, hence the field equations become easy to solve and the stress-energy tensor for this system satisfies the perfect fluid description of matter \cite{Madsen}. 

Now, we apply these transformation rules to some known geometries and obtain the expansion rates for each spacetime generated by this method. Therefore, we shall see that the vacuum energy provides the expansion of the universe to accelerate, and the matter content causes it to decelerate in time.

Our paper is organised as follows.  In section 2, we study on the Friedmann-Lemaitre-Robertson-Walker geometry and write the line element in the conformal form. First we apply time dependent conformal factor acting on Minkowski geometry and we obtain the expansion rates as a function of scalar field. Next, we consider that the conformal factor is acting on a curved static spacetime, and obtain that expansion rate decreases due to curvature of the submanifold. In Section 3, we introduce a conformally rescaled, charged, rotating Kerr Black hole  with time dependent conformal factor and we obtain a relation between scale factor and BD scalar field. Finally, we end the paper with a brief summary and concluding remarks.

\section{Friedmann-Lemaitre-Robertson-Walker cosmological solution}

FLRW line element is a well known cosmological metric and given by,
\begin{equation}
	d\tilde{s}^2=-dt^2+a(t)^2\left(\frac{dr^2}{1-kr^2}+r^2d\Omega^2\right),
\end{equation} 
here, $t$ is a cosmological time and $a(t)$ is the scale factor, $k$ is the spatial curvature parameter and $d\Omega^2$ is two sphere metric. If we rescale the cosmological time as $dt=S(\tau)d\tau$ and reorganize the metric suitably, we can write FLRW line element in the conformal form,
\begin{eqnarray}
	d\tilde{s}^2&=&S(\tau)^2ds^2,\nonumber\\
	\nonumber\\
	&=&S(\tau)^2\left[-d\tau^2+\frac{dr^2}{1-kr^2}+r^2d\Omega^2\right] ,
\end{eqnarray}
by denoting $\tau $ as a conformal time and $S(\tau)^2$ as the conformal factor where ${a\left(t\right)}^2={S[\tau \left(t\right)]}^2$. Therefore, the static part in square brackets, which we call as the submanifold, is transformed to the cosmological spacetime by a time dependent conformal factor $S(\tau)^2$. The Ricci scalar of the cosmological space-time is
\begin{equation}
	\tilde{R}=\frac{6}{S^2}\left[k+\frac{\ddot{S}}{S}\right],
\end{equation}
where the overdot represents derivative with respect to conformal time $\tau $. The Ricci scalar of submanifold reads $R=6\,k$, which will be vanished for $k=0$, hence, we obtain a Minkowski line element that implies the flat  submanifold.  The BD solution for FLRW metric has been explicitly given by \cite{Bergh,Weinberg,Cervero}. Also, the work \cite{Ciftci1} reviews all possible solutions in this subject. Strictly speaking that, the main purpose of our work is not to obtain all the solutions and repeat the literature but try to understand the effect of the scalar field for some time dependent spacetimes with static or stationary submanifold and compare their expansion rates in the subject of stress-energy tensor for the perfect fluid.

Using the and equation \eqref{perfectfluid} we solve the BD field equation \eqref{SEtensor} through the computer algebra and obtain that,
\begin{eqnarray}
	\frac{8\pi S{\left(\tau \right)}^2}{\tilde{\phi }\left(\tau \right)}\tilde{\rho }\left(\tau \right)&=&3k+\frac{3{\dot{S}(\tau )}^2}{{S(\tau )}^2}+\frac{3\dot{S}(\tau )\dot{\phi }(\tau )}{S(\tau )\phi (\tau )}-\frac{\omega }{2}\frac{{\dot{\phi }\left(\tau \right)}^2}{{\phi \left(\tau \right)}^2} \label{BDeq1} \\
	\frac{8\pi S{\left(\tau \right)}^2}{\widetilde{\phi }\left(\tau \right)}\tilde{P}\left(\tau \right)&=&-k+\frac{{\dot{S}\left(\tau \right)}^2}{{S\left(\tau \right)}^2}-\frac{2\ddot{S}\left(\tau \right)}{S\left(\tau \right)}-\frac{\dot{S}\left(\tau \right)\dot{\phi }\left(\tau \right)}{S\left(\tau \right)\phi \left(\tau \right)}-\frac{\omega }{2}\frac{{\dot{\phi }\left(\tau \right)}^2}{{\phi \left(\tau \right)}^2}-\frac{\ddot{\phi }\left(\tau \right)}{\phi \left(\tau \right)}\label{BDeq2}\,,
\end{eqnarray}
and using the ansatz (\ref{phi&S}), we obtain the energy density $\tilde{\rho }$ and pressure $\tilde{P}$ as,
\begin{eqnarray}
	\tilde{\rho}&=&\frac{S^{\alpha-2}}{16\pi}\left(6k+(6+6\alpha-\omega\alpha^2)\frac{\dot{S}^2}{S^2}\right)\, ,\\
	\tilde{P}&=&-\frac{S^{\alpha-2}}{16\pi}\left[2k+((2+\omega)\alpha^2-2)\frac{\dot{S}^2}{S^2}+2(2+\alpha)\frac{\ddot{S}}{S}\right]\, ,
\end{eqnarray} where the four velocity vector has only time component as $\tilde{u}_a=\{-S(\tau) ,0,0,0\}$. Substituting $\tilde{\phi}=\phi_0\,S(\tau)^{\alpha }$, the scalar field equation (\ref{Scalarfield}) satisfies,
\begin{equation}
	6k-\omega\alpha(2+\alpha)\frac{\dot{S}^2}{S^2}+2(3-\omega\alpha)\frac{\ddot{S}}{S}=0.\label{key}
\end{equation}
Here, the energy conservation equation $\tilde{\nabla }^a{\tilde{T}}^{\,\,\,(pf)}_{ab}=0$ becomes identical with the \eqref{key} and satisfies the same equation. The (\ref{key}) is a key equation and responds to the question of how the flat subspace yields expansion of the spacetime to be accelerated or how any type of the content of this subspace affects the expansion rate of the universe.

\textbf{If} $\mathbf{k=0}$; The untransformed submanifold has no curvature and becomes Minkowski line element, thus it contains no matter, therefore, solving (\ref{key}), this flat submanifold yields conformal factor to be a power law,
\begin{equation}
	S(\tau)=S_0\,  \tau^{\frac{2\omega\alpha-6}{\omega\alpha(\alpha+4)-6}}\label{conformalfactor}\, ,
\end{equation}
where, $S_0$ is an integration constant Therefore, the empty submanifold is transformed to the cosmological vacuum era by a conformal factor. If we rescale the time parameter as $S(\tau) d\tau=dt$ and rewrite the scale factor and scalar field in terms of the cosmological time, we get
\begin{equation}
	a(t)=a_0\,  t^{\frac{2\omega\alpha-6}{\omega\alpha(\alpha+6)-12}}\quad \mbox{and} \quad \tilde{\phi}(t)=\phi_0\, a(t)^\alpha \label{Scaleparameter},
\end{equation}
where the constant, $a_0={S_0}^{\frac{\omega \alpha \mathrm{(}\alpha \mathrm{+4)-6}}{\omega \alpha \mathrm{(}\alpha \mathrm{+6)-12}}}{\left(\frac{\omega \alpha (\alpha +6)\mathrm{-}\mathrm{12}}{\omega \alpha \mathrm{(}\alpha \mathrm{+4)-6}}\right)}^{\frac{\mathrm{2}\omega \alpha \mathrm{-}\mathrm{6}}{\omega \alpha \mathrm{(}\alpha \mathrm{+6)-12}}}$ . These solutions are consistent with previous results \cite{Ciftci1}. Here, note that if the power of $\tau $, in \eqref{conformalfactor}, is equal to $-1$, the scale factor in \eqref{Scaleparameter} will be an exponential function, hence we obtain de-Sitter spacetime. Nevertheless, this result is a very special subcase of the solution presented in our work, and we prefer to stay in the power-law type solution. Another motivation to insist on this solution group is to keep the whole paper in the same context. That means we aim to compare the expansion rates of three differently curved spacetimes and the common properties of these geometries are all admit the power-law expansion parameter simultaneously.

Based on this setup, the deceleration parameter that gives how the universe accelerates, becomes
\begin{equation}
	q=-\frac{\ddot{a}a}{\dot{a}^2}=\frac{\omega\alpha^2+4\alpha-6}{2\omega\alpha-6},
\end{equation} 
which strongly depends on the relation between the expansion parameter and the scalar field. In this result, $\alpha $ remains as a free parameter and we can determine the value of $\alpha $ from cosmological observations. Depending on the value of $\alpha $, we may have acceleration or deceleration of the spacetime.

\textbf{If} $\mathbf{k\neq 0}$ : This choice describes a submanifold with a constant curvature and causes a nonzero stress-energy tensor. Hence, the first term $G_{ab}$ on the right side of \eqref{SEtensor} has some contribution to the system. Physically, that means we are studying a homogeneously curved submanifold, and hence this submanifold has some massive content and affects the expansion rate.

The solution of nonlinear differential equation \eqref{key} has the form of an exponential equation,
\begin{equation}
	S(\tau)= \,S_0\, e^{\pm\left(\frac{6k}{\omega \alpha^2+4\omega\alpha-6}\right)^{1/2}\tau},
\end{equation}
and the expansion parameter with respect to cosmological time becomes,
\begin{equation}
	a(t)=	\pm \left(\frac{6k}{\omega\alpha^2+4\omega\alpha-6}\right)^{1/2}t\quad \mbox{and}\quad \tilde{\phi}(t)=\phi_0\, a(t)^\alpha,
\end{equation}
which satisfies a linearly expanding spacetime and fits the result obtained in \cite{Ciftci1}. Here, note that we choose the plus sign for the consistency. This result could be interpreted as follows, the matter content in the conformally transformed submanifold prevents the acceleration of the spacetime, or we may say that the scalar field in the curved region could not accelerate the expansion of the spacetime. Neverheless, the vacuum submanifold that fills with a scalar field could speed up the expansion of spacetime. In summary, by taking into account $k=0$ and $k{\neq }0$ cases, we can propose the following statements: while a scalar field yields an accelerated expansion for the flat submanifold, it cannot accelerate the curved manifold filled with matter. This result might be applied to cosmology and interpreted as: a galactic system does not expand locally however, the vacuum parts of the universe are expanding much more and spreading apart the galaxies from each other.

\section{The  Charged and Rotating Time Dependent Black Hole Solution}

In this part, using the same ansatz in previous part, we search for an allowable cosmological black hole solution. A cosmological black hole might be possible by means of embedding a static or stationary black hole in a cosmological background. There are some similar cosmological black hole geometries considered in the literature \cite{Thakurta,Sultana,McClure,Moradpour,Hansraj,McVittie}. More realistic black holes are axially symmetric ones that have a mass and angular momentum. Although it is not necessary to have electrical charge for the physically reasonable black holes, to obtain a more general result, we include the electrical charge in this work. Therefore, we consider a Kerr-Newman (KN) metric that will be transformed into a spacetime varying in time with a time-dependent conformal factor. Since observations show that the realistic black holes curve the spacetime around themself, and also they are dynamical objects and interact with their environment, it will be convenient to work with this geometry that changes in time. From the previous part, we expect that, these massive objects, in which we may call the curved stationary submanifolds, might decelerate the expansion rate around itself or cause spacetime to be expanded linearly or there might be no expansion at all.

The simplest way to embed a black hole in a time dependent framework is to multiply all $\mathrm{KN}$ metric by a time dependent scale factor $S\mathrm{(}\tau {\mathrm{)}}^{\mathrm{2}}$ \cite{Sultana}. Then the metric becomes,
\begin{eqnarray}
	\tilde{ds}^2&=&S(\tau)^2\,ds^2_{\, KN}\label{Kerrmetric} \, ,\\&&\nonumber\\&=&S(\tau)^2\Big[-\Big(1-\frac{2Mr-Q^2}{\Sigma}\Big)d\tau^2-2a\sin^2\theta\frac{(2Mr-Q^2)}{\Sigma}d\tau d\varphi\nonumber\\&&\nonumber\\&&+\Sigma\Big(\frac{dr^2}{\Delta}+ d\theta^2\Big) +\Big(a^2+r^2+\frac{(2Mr-Q^2)a^2\sin^2\theta}{\Sigma}\Big)\sin^2\theta d\varphi^2\Big]\, ,\nonumber
\end{eqnarray}
which can also be written as,
\begin{eqnarray}
	\tilde{ds}^2&=&S(\tau)^2\Big[-d\tau^2+\Sigma\Big(\frac{dr^2}{\Delta}+ d\theta^2\Big)+\frac{2Mr-Q^2}{\Sigma}\left(-d\tau+a\sin^2\theta d\varphi\right)^2\nonumber\\&&\nonumber\\&&+(a^2+r^2)\sin^2\theta d\varphi^2\Big]\, ,\label{Kerrmetric2}
\end{eqnarray}
where,
\begin{eqnarray}
	\Sigma&=&r^2+a^2\cos^2\theta\\
	\Delta&=&r^2+a^2+Q^2-2Mr,
\end{eqnarray}
where, $M$ is the mass, $Q$ is the electric charge and $a$ is the the rotation parameter of the body. Here, for large radial distances, this geometry reduces to the spatially flat FLRW geometry. For the systems with electromagnetic field, the Brans-Dicke action is given by
\begin{equation}
	\tilde{S}_{BD}= \int d^4x \sqrt{ -\tilde{g}} \left\{ \tilde{\phi} \tilde{R} 
	-\omega \frac{ \tilde{g}^{ab}}{\tilde{\phi}} \tilde{\nabla}_a \tilde{\phi} 
	\tilde{\nabla}_b \tilde{\phi}-\tilde{F}^{ab}\tilde{F}_{ab}+\tilde{\mathcal{L}}_m \right] \,,\label{BDaction}
\end{equation}
where, $\tilde{R}$ is the Ricci scalar of overall cosmological metric, $\tilde{\phi }$ is the BD scalar field, $\omega $ is the BD parameter, ${\tilde{F}}^{ab}$ is the Maxwell electromagnetic tensor and ${\tilde{\mathcal{L}}}_m$ is the Lagrangian density for the matter part. The total stress-energy tensor for this cosmological background contains electromagnetic and perfect fluid contributions as
\begin{eqnarray}
	\tilde{T}_{ab}&=&\tilde{T}_{ab}^{\,(EM)}+\tilde{T}_{ab}^{\,(m)}\, ,\nonumber\\
	&&\nonumber\\
	&=&\frac{T_{ab}^{\,(EM)}}{S(\tau)^2}+\tilde{T}_{ab}^{\,(pf)}\label{SETensorKerr}.
\end{eqnarray}
The Ricci scalar for this geometry becomes
\begin{equation}
	\tilde{R}=\frac{6[\Delta \Sigma-(r^2+a^2)(Q^2-2Mr)]}{\Delta \Sigma}\frac{\ddot{S}}{S^3}=\frac{6\ddot{S}}{S}|g^{00}|\label{RicciKN}.
\end{equation}
Here, the matter part might be chosen as the form given in \eqref{perfectfluid} and the velocity four vector has the following components:
\begin{equation}
	\tilde{u}_a=\{-S(\tau)\sqrt{\frac{\Delta \Sigma}{\Delta \Sigma-(r^2+a^2)(Q^2-2Mr)}},0,0,0\}=\{-\frac{1}{\sqrt{|\tilde{g}^{\,00}|}},0,0,0\}\, ,
\end{equation}
and ${\tilde{u}}_a{\tilde{u}}^a=-1$.

The nonzero components of energy-momentum tensor for the matter part are
\begin{eqnarray}
	\tilde{T}^\tau_{\tau}\,^{(pf)}&=&-\tilde{\rho}(\tau,r,\theta)\, ,\nonumber\\
	&&\nonumber\\
	\tilde{T}^r_{\ r}\,^{(pf)}&=&\tilde{T}^\theta_{\ \theta}\,^{(pf)}=\tilde{T}^\varphi_{\ \varphi}\,^{(pf)}=\tilde{P}(\tau,r,\theta)\, ,\nonumber\\
	&&\nonumber\\
	\tilde{T}^{\varphi}_{\tau}\,^{(pf)}&=&\frac{2a(Q^2-2Mr)(\tilde{\rho}+\tilde{P})}{\Delta\Sigma-(r^2+a^2)(Q^2-2Mr)}\, .
\end{eqnarray}
The electromagnetic stress-energy momentum tensor is given by
\begin{equation}
	\tilde{T}_{ab}^{(EM)}=2(\tilde{F}_{ac}\tilde{F}_{bd}g^{cd}-\frac{1}{4}\tilde{F}_{cd}\tilde{F}^{cd}g_{ab}).\label{TEM}
\end{equation} Here, the electromagnetic potential one form is,
\begin{equation}
	A_a=\Big(-\frac{Qr}{\Sigma},0,0,\frac{Qra\sin^2\theta}{\Sigma}),
\end{equation}
and the electromagnetic field tensor is given by $\tilde{F}=\tilde{d}A$ or in component form, it is defined as $\tilde{F}_{ab}=\tilde{\nabla}_a A_b- \tilde{\nabla}_b A_a$. The nonzero components of electromagnetic energy-momentum tensor are
\begin{eqnarray}
	\tilde{T}^\tau_{\ \tau}\,^{(EM)}&=&-\tilde{T}^\varphi_{\ \varphi}\,^{(EM)}=\frac{Q^2[\Sigma-2(r^2+a^2)]}{\Sigma^3S^4},\nonumber\\
	&&\nonumber\\
	\tilde{T}^\tau_{\ \varphi}\,^{(EM)}&=&\frac{2aQ^2(r^2+a^2)\sin^2\theta}{\Sigma^3 S^4},\nonumber\\
	&&\nonumber\\
	\tilde{T}^r_{\ r}\,^{(EM)}&=&T^\theta_{\ \theta}\,^{(EM)}=\frac{Q^2}{2\Sigma^2 S^4},\nonumber\\
	&&\nonumber\\
	\tilde{T}^\varphi_{\ \tau}\,^{(EM)}&=&\frac{2aQ^2}{\Sigma^3S^4}\, .
\end{eqnarray}
Note that, the conservation of energy-momentum tensor for the electromagnetic part, ${\tilde{\mathrm{\nabla }}}^a{\tilde{T}}^{\mathrm{(}EM\mathrm{)}}_{ab}\mathrm{=}$ 0 is already satisfied.

Substituting the stress-energy tensor \eqref{SETensorKerr} in the Brans-Dicke field equations \eqref{SEtensor} and using computer algebra,  from the $(\tau,r)$ term, we obtain the following differential equation,
\begin{equation}
	2\tilde{\phi}(\tau)\dot{S(\tau)}+\dot{\tilde{\phi}}(\tau)S(\tau)=0,
\end{equation}
and the solution for the scalar field is given by,
\begin{equation}
	\tilde{\phi}(\tau)=\phi_0\, S(\tau)^{-2}\label{P1}.
\end{equation}
The components $(\tau,\tau)$ and $(r,r)$ of the field equations \eqref{SEtensor} are satisfied for $\phi_0=1$ then the energy density and pressure become,
\begin{equation}
	\tilde{\rho}(\tau,r,\theta)=\tilde{P}(\tau,r,\theta)=-\frac{(2\omega+3)\Big(\Delta\Sigma-(Q^2-2Mr)(r^2+a^2)\Big)}{8\pi\Delta\Sigma}\frac{\dot{S}^2}{S^6}\, .\label{energydensity}
\end{equation}
To satisfy positive energy density, this result requires to be $\omega <-\frac{3}{2}$ or $\Delta\Sigma<(Q^2-2Mr)(r^2+a^2)$. In the works \cite{Sultana,McClure}, energy density becomes negative for the cosmological black hole geometries generated in this way. Therefore, using a straightforward conformal transformation, in BD theory, we have embedded a Kerr-Newman black hole in an expanding universe filled with matter. The equation of state, $\tilde{\rho }=\tilde{P}$ is known as Zeldovich's stiff fluid model and is used in general relativity to obtain the stellar and cosmological models for ultrahigh dense matter \cite{Zeldovich}.

Now, we must satisfy the scalar field equation (\ref{Scalarfield}),
\begin{equation}
	\frac{2(2\omega+3)\Big(\Delta\Sigma-(Q^2-2Mr)(r^2+a^2)\Big)}{\Delta \Sigma }\frac{\ddot{S}}{S^5}=0\,.\label{BDfe3}
\end{equation} 
This equation restricts the scale factor $S(\tau)$  to be linearly depending on time as,
\begin{equation}
	S(\tau)=S_0\, \tau.\label{scalefactor2}
\end{equation}
This value of scale factor also satisfies the conservation of energy-momentum tensor for the matter part and given by,
\begin{eqnarray}
	\tilde{\nabla}^a\tilde{T}_{ab}^{\,(pf)}=0\label{BDfe5}.
\end{eqnarray} 
Therefore, all of the field equations are satisfied and the line element takes the folowing form,
\begin{eqnarray}
	ds^2&=&{S_0}^2 \tau^2\Big[-\Big(1-\frac{2Mr-Q^2}{\Sigma}\Big)d\tau^2-2a\sin^2\theta\frac{(2Mr-Q^2)}{\Sigma}d\tau d\varphi\nonumber\\&&\nonumber\\&&+\Sigma\Big(\frac{dr^2}{\Delta}+ d\theta^2\Big)+\Big(a^2+r^2+\frac{(2Mr-Q^2)a^2\sin^2\theta}{\Sigma}\Big)\sin^2\theta d\varphi^2\Big]. \nonumber\\
	\label{Kerrmetric11}
\end{eqnarray}
To analyse the singularity structure of this spacetime, since Ricci scalar \eqref{RicciKN} is zero for the result \eqref{scalefactor2}, we can look for the square of Riemann tensor
\begin{eqnarray}
	\tilde{R}_{abcd}\tilde{R}^{abcd}=\frac{f(r,\theta)}{\tau^8\Delta^2 \Sigma^4},
\end{eqnarray}
where the function, $f(r,\theta)$ in the numerator is an $r$ and $\theta $ dependent complicated function and its explicit expression is not needed to determine the singularity structure of the geometry. This geometry posesses three singular points, namely, the initial big bang type singularity at $\tau=0$, the ring singularity of Kerr-Newman solution at $\Sigma=0$, and also the horizon singularity at $\Delta=0$. These singularities also cause a singular fluid in which the energy density and pressure (\ref{energydensity}) diverge as well. If we rescale the conformal time, the line element (\ref{Kerrmetric11}) can be expressed as,
\begin{eqnarray}
	ds^2&=&-\Big(1-\frac{2Mr-Q^2}{\Sigma}\Big)dt^2+t\Big[-2a\sin^2\theta\frac{(2Mr-Q^2)}{\sqrt{t}\Sigma}dt d\varphi\nonumber\\&&\nonumber\\&&+\Sigma\Big(\frac{dr^2}{\Delta}+ d\theta^2\Big)+\Big(a^2+r^2+\frac{(2Mr-Q^2)a^2\sin^2\theta}{\Sigma}\Big)\sin^2\theta d\varphi^2\Big].\nonumber\\ \label{Kerrmetric12}
\end{eqnarray}
Here, the scale factor has the form $a\mathrm{(}t\mathrm{)=}\sqrt{t}$ hence, the deceleration parameter becomes $q=1$ (where the integration constants are chosen to be unity as a convenience). This result shows that the spacetime around a charged rotating object with mass $M$ and angular momentum $a$ is not accelerating but decelerating as we expect. The scalar field in this curved submanifold does not yield the expansion to accelerate.
\section{Conclusion}

In this work, we have tried to explain how the vacuum energy provides the expansion of the universe to be accelerated in time and how the matter content of spacetime causes the universe to be decelerated. Using the rules for conformal transformation of a metric and the BD theory, we get some different cosmological spacetimes from the several static or stationary submanifolds. One of these submanifolds has chosen as Minkowskian spacetime, one has constant curvature, and the other has a massive content. The conformal factor has been set as a time dependent function, and the Brans-Dicke scalar field is directly related to this conformal factor as $\tilde{\phi }=\phi_0\, S(\tau)^{\alpha }$. Depending on the matter content in the submanifold, we obtain different expansion rates resulting in various scalar fields for each scenario. We conclude that the BD scalar field yields an accelerated expansion for the empty submanifold, but it is difficult to expand a spacetime filled with some pressure and energy. Therefore, the scalar field becomes responsible for the expansion in the vacuum. On the other hand, the gravitational sector prevents the expansion of spacetime even if there is a scalar field existing around the massive content. Cosmologically, the effect of scalar field can be explained as follows: If a spacetime has some massive content in it, this spacetime is not expanded so fast, nevertheless, an empty spacetime can have accelerated expansion due to the BD scalar field that might be interpreted as the effect of dark energy in the conventional cosmology.

\section*{Acknowledgement}

We are grateful to V. Faraoni for the discussion, and many thanks to O. Delice for his valuable comments. Also we  thank Tekirdağ Namık Kemal University.

\end{document}